\documentclass[aps,prb,twocolumn, unsortedaddress, superscriptaddress, showpacs]{revtex4}
\usepackage{amsmath,amsthm}
\usepackage{epsfig}
\usepackage{graphicx}
\usepackage{dcolumn}
\usepackage{bm}
\usepackage{hyperref}

\begin{document}

\title{Optical probe of strong correlations in LaNiO$_{3}  $ thin films}
\author{M. K. Stewart}
\email{mstewart@physics.ucsd.edu}
\affiliation{Department of Physics, University of California-San Diego, La Jolla, California 92093, USA} 
\author{ Jian Liu}
\affiliation{Department of Physics, University of Arkansas, Fayetteville, Arkansas 72701, USA}
\author{  R. K. Smith}
\author{ B. C. Chapler}
\affiliation{Department of Physics, University of California-San Diego, La Jolla, California 92093, USA} 
\author{ C.- H. Yee} 
\author{K. Haule}
\affiliation{Department of Physics $\&$ Astronomy, Rutgers University, Piscataway, New Jersey 08854-8019, USA}
\author{ J. Chakhalian}
\affiliation{Department of Physics, University of Arkansas, Fayetteville, Arkansas 72701, USA}
\author{D. N. Basov}
\affiliation{Department of Physics, University of California-San Diego, La Jolla, California 92093, USA}

\date{\today}

\begin{abstract}

The optical properties of LaNiO$_3$ thin films are investigated over a wide energy and temperature range.  Thin films of varying thickness were epitaxially grown by pulsed laser deposition on LaAlO$_3$ and SrTiO$_3$ substrates.  The optical conductivity data of the films reveal a number of interband transitions above 1 eV which are in good agreement with band structure calculations.  No well defined Drude peak is observed however, in stark contrast with LDA theory predicting a finite density of states at the Fermi energy.  This experimental finding of a vanishing Drude spectral weight, compared to a finite electron kinetic energy obtained from band structure calculations, highlights the importance of strong electronic correlations in LaNiO$_3$.

\end{abstract}


\maketitle

\section{Introduction}

Transition metal oxides exhibit a wide array of interesting physical phenomena such as high temperature superconductivity, insulator-to-metal transitions, and half-metallic ferromagnetism.\cite{goodenough}  Modern sample growth technologies offer an opportunity to control and tune these effects through heterostructuring.\cite{PLD_growth}  At the interface between two oxides with different properties, the competing order parameters can dramatically modify the orbital, electronic, and magnetic structure of the bulk materials.\cite{science_interfaces}  Switching of superconductivity on and off, for instance, has been demonstrated at the interface between an oxide superconductor and an oxide ferroelectric.\cite{superconductivity_tuning}  New properties not present in the constituent materials can also emerge at these interfaces. The formation of a metallic layer at the interface between the insulators LaAlO$_3$ and SrTiO$_3$, for example, has already been demonstrated.\cite{LAO_STO}  Because of the ample choice of exotic properties offered by transition metal oxides for the growth of superlattices, experimental efforts will greatly benefit from the guidance of theoretical work.  It is therefore of great importance that the constituent materials be thoroughly characterized experimentally and that their properties be well understood theoretically.  This can be especially challenging, given that strong electronic correlations are often present in transition metal oxides.\cite{tokura_phys_today}

One of the most ambitious heterostructuring ideas yet is that of superconducting LaNiO$_3$/LaAlO$_3$ superlattices.\cite{superlattice_theory_first, superconductivity_theory}  Since the discovery of high-temperature superconductivity in the cuprates, the search for new superconducting materials with potentially higher transition temperatures has been at the vanguard of condensed matter Physics research.  LaNiO$_3$ (LNO) is in many ways similar to the cuprates, but with one important difference: it has one electron in two degenerate, three-dimensional \textit{e$_{g}$} orbitals.  The lack of orbital degeneracy and the quasi two-dimensionality of the \textit{e$_{g}$} electrons are well established attributes of superconductivity in the cuprates.  Modifying the orbital structure of LNO through layered heterostructuring, and successfully inducing superconductivity in this non-superconducting material, would be a major breakthrough.  Because of this, and given our limited understanding of LNO thus far, a detailed optical study of this oxide promises to be of great use.

In the bulk, LNO is a paramagnetic metal with a rhombohedrally distorted perovskite crystal structure.\cite{structure}  Resistivity, susceptibility and heat capacity data reported for powder LNO are suggestive of strong correlations in this system.\cite{correlations_transport}  Unlike other rare earth nickelates, LNO does not exhibit a temperature controlled metal-insulator transition, but an antiferromagnetic insulating state has been reported in oxygen deficient LaNiO$_{3-x}$.\cite{oxygen_MIT}  In this work we show that LNO is a correlated electron system, as evidenced by an experimental Drude spectral weight that is strongly suppressed compared to that predicted by band structure calculations.  The optical data presented here can not be fully described by available theoretical models of LNO.  We propose that a better theoretical understanding of LNO is essential in order to fully explore the possibilities and limitations of LaNiO$_3$/LaAlO$_3$ superlattices.

\section{Experimental Methods}

Bulk synthesis of LNO  has been a challenging issue, resulting in available single crystals limited to the micron size. In sharp contrast, depositing epitaxial films with relatively low oxygen pressure is largely possible because of the epitaxial stabilization.\cite{strain} Thus, by pulsed laser deposition, we grew epitaxial LNO films with thicknesses that are sufficiently larger than the critical thickness, therefore minimizing strain-induced distortions. To further isolate the effect of strain, 001-oriented LaAlO$_3$ (LAO) and  SrTiO$_3$ (STO) single crystal substrates with -1.2$\%$ and +1.7$\%$ lattice missmatch, respectively, were used for comparison. X-ray diffraction showed that the average c-axis parameters are 3.84~\AA\ and 3.86~\AA\ on STO and LAO, respectively. Small amounts (limited to a few percent) of secondary phases, such as NiO, were also detected due to the low thermodynamic stability of rear-earth nickelates and the decreasing epitaxial stabilization with increasing thickness.\cite{strain}  Transport data of the films was obtained between 2~K and 400~K using a four-point probe method in the \textit{van der Pauw} geometry.  

Optical studies of both the films and the bare substrates were carried out using reflectance in the range from 50 to 700 cm$^{-1} $ and variable angle spectroscopic ellipsometry (VASE) in the range from 700 to 48000 cm$^{-1} $.  Near-normal incidence reflectance measurements were performed in a Michelson interferometer (Bruker 66vs).  Reflectance of the sample was first measured relative to a gold reference mirror and then normalized by the reflectance of the gold coated sample.\cite{Homes_Au}  Ellipsometry measurements were performed with two commercial Woollam ellipsometers.  The range from 700 to 4500 cm$^{-1} $ was investigated with an IR-VASE model based on a Bruker 66vs.  For the range between 5000 and 48000  cm$^{-1} $ we used a VASE model based on a grating monochromator.  Both ellipsometers are equipped with home-built UHV chambers to allow low temperature measurements.\cite{Ken_UHVellip}  All the samples were characterized over the entire frequency range at room temperature and in some cases at low temperatures down to 20K.  Ellipsometry measurements were performed at incidence angles of 60$^{\circ} $ and 75$^{\circ} $.  At each angle, the polarization state of the reflected light was measured in the form of two parameters, $ \Psi$ and $\Delta$, which are related to the Fresnel reflection coefficients for $\textit{p-}$ and $\textit{s-}$ polarized light (\textit{R$_{p}$} and \textit{R$_{s}$}) through the equation $\frac{R_p}{R_s}=tan(\Psi)e^{i\Delta}$.  

In order to obtain the optical constants from the raw reflectance and ellipsometry data, a model was created using Kramers-Kronig consistent Lorentz oscillators to describe the complex dielectric function of the sample.  The parameters in the model were then fitted to the experimental data using regression analysis.\cite{ellipsometry}  In the case of the LNO films, the model consisted of two layers: a substrate characterized by the optical constants previously determined for either LAO or STO, and a thin film layer from which the optical constants of the film alone were obtained.

\begin{figure}[htb]
\centering
\includegraphics[width=86mm]{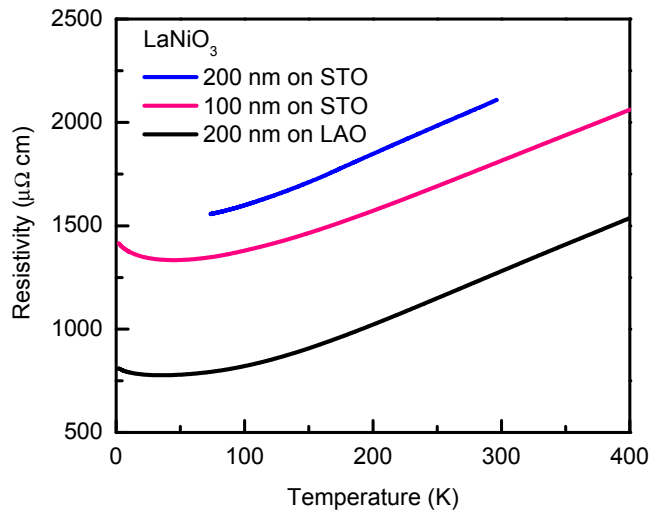}
\caption{(color online) DC resistivity of the LNO films over a wide temperature range.} \label{transport}
\end{figure}

\section{Results and Discussion}

Transport data shown in Fig~\ref{transport} reveal the resistivity of the films to be on the order of 1 m$\Omega$ cm.  This is in reasonable agreement with previously reported values.\cite{correlations_transport, transport}   The data show metallic behavior of the samples, with only a small temperature dependence.  The 200 nm film on STO shows resistivity a factor of two higher than that on LAO throughout the entire temperature range.  Comparing the 100 nm and 200 nm films on STO, the thinner film exhibits a resistivity about 15$\%$ lower than the thicker one.

\begin{figure}[htb]
\centering
\includegraphics[width=86mm]{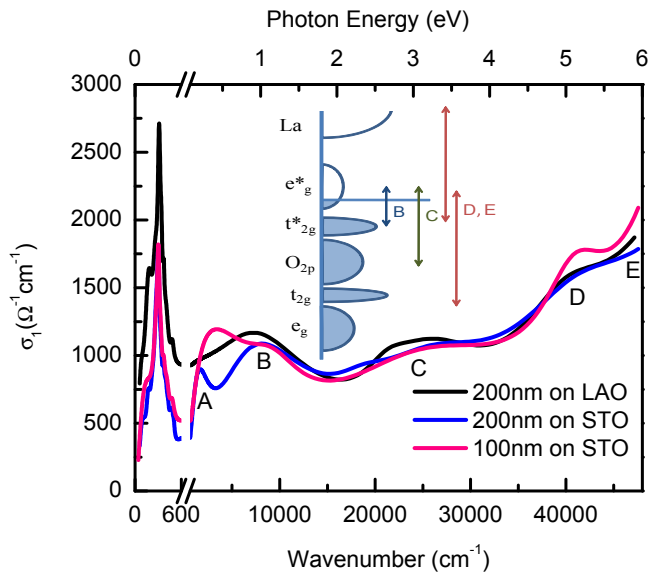}
\caption{(color online) Real part of the optical conductivity of three different LNO films at room temperature.  Note the change in scale at 600 cm$^{-1}$.  Inset: rough sketch of the LNO density of states and interband transitions.} \label{sigma1_roomT}
\end{figure}

To gain insight into the electronic band structure of LNO it is useful to look at the optical conductivity, which is related to the complex dielectric function through the equation $ \sigma(\omega)=\frac{i\omega[1-\epsilon (\omega)]}{4\pi} $.  Figure~\ref{sigma1_roomT}  shows the real, dissipative part of the optical conductivity at room temperature.  In the high energy region, above 1 eV, the main features of the optical conductivity, which we have labelled B-E, look similar for all the films.  A more noticeable variation in the spectra is evident in the mid-IR region, presumably due to the effects of strain.  While all samples show a peak in this region (A in Fig.\ref{sigma1_roomT}), the magnitude and shape is different for each film.  In the far-IR range (note the change in scale), we see a broad peak centred at 300 cm$^{-1}$.  In stark contrast with theoretical predictions, no well defined Drude peak is observed, indicative of very strong electronic correlations.  Below we discuss in detail the data in these three regions.  The similarity of 100nm and 200nm films on STO seen in Fig.\ref{sigma1_roomT}, especially in the far-IR region, shows that the secondary phases that may develop with increasing thickness do not significantly affect the optical conductivity of the films.  Additionally, the similarity of the films on STO and LAO indicates that our data indeed reflect the properties of strain-free LNO.

\begin{figure}[htb]
\centering
\includegraphics[width=86mm]{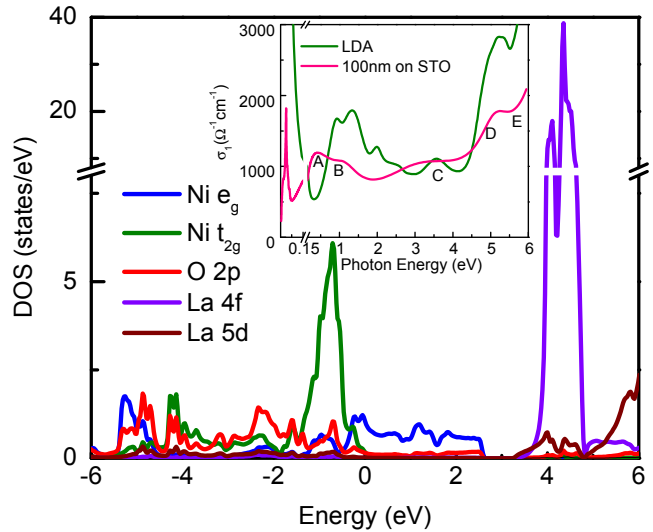}
\caption{(color online) Partial LNO density of states obtained from LDA theory.  Inset: comparison of the optical conductivity obtained experimentally for the 200 nm film of LNO on STO and that obtained from LDA calculations.  The Drude peak produced by LDA theory is not present in the experimental data.} \label{theory}
\end{figure}

\begin{figure}[htb]
\centering
\includegraphics[width=86mm]{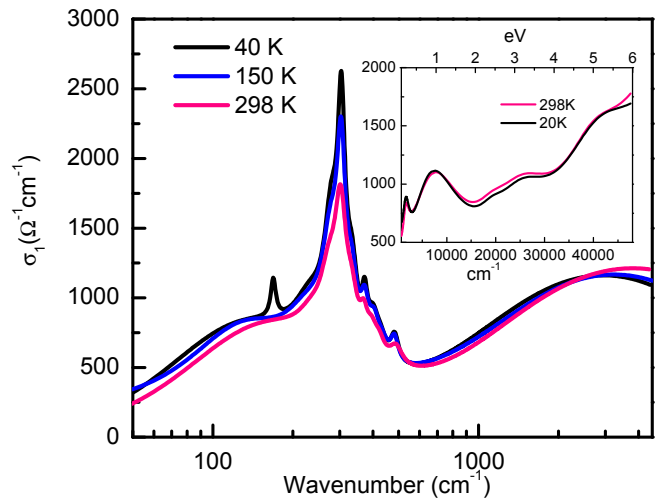}
\caption{(color online) Low temperature optical conductivity of the 100 nm thick film on STO in the far-IR and mid-IR regions.  Inset: optical conductivity of the 200 nm film on STO up to 6 eV at 298 and 20 K.} \label{sigma1_lowT}
\end{figure}

The high energy optical conductivity we have acquired is consistent with  earlier work on LNO ceramic samples.\cite{tokura_prb, Tokura_jpsj}  The present LDA band structure was computed within the  FP-LAPW scheme~\cite{dmft} using room temperature bulk LNO structural parameters,\cite{structure} and is in good agreement with earlier bandstructure calculations.\cite{Sarma_band_structure, sigma1_theory, Nohara_band_structure}  In Fig.\ref{theory}, we plot the calculated partial density of states, which indicates LNO has a $t_{2g}^{6}e_{g}^{1}$ electronic configuration, with the antibonding \textit{e$_{g}$} states crossing the Fermi level.  Based on the data in Fig.\ref{theory} and using transition decomposition analysis of the LDA optical conductivity, we attempt to assign features B-D in Fig.\ref{sigma1_roomT} to specific interband transitions (see inset in Fig.\ref{sigma1_roomT}).  We suggest that B corresponds to transitions from the Ni \textit{t*$_{2g}$} and \textit{e*$_{g}$} levels to the Ni \textit{e*$_{g}$} orbitals .  C is due to transitions from the O 2\textit{p} to the \textit{e*$_{g}$} orbitals.  D and E are the result of transitions from \textit{t*$_{2g}$} to the La 4f and 5d levels and from the bonding Ni \textit{e$_{g}$} and \textit{t$_{2g}$} orbitals to the \textit{e*$_{g}$} orbitals.  Low temperature optical conductivity in this range is shown in the inset of Fig.\ref{sigma1_lowT}.  All of the features remain unchanged down to 20 K.  A small reduction of up to 5 $\%$ in the optical conductivity is evident between 1.5 and 4 eV.  

Based on LDA calculations, feature A in the mid-IR region appears to be too low in energy to be an interband transition.  However, given the failure of this model to accurately describe our low frequency data, it is possible that the \textit{e*$_{g}$} band is, in reality, significantly different from that obtained by these calculations.  In this scenario, it can not be ruled out that this peak is indeed associated with an interband transition.

Alternatively, feature A could be a sign of some kind of electron localization.  It is known that localization phenomena can give rise to a shift of the Drude peak to finite frequencies, due to the need of the electrons to overcome an energy barrier.\cite{Dimitri_rmp}   We note that the peak does seem to be centred at an unusually high energy compared to other materials where localization of electrons is present.\cite{Dimitri_Y123, Dimitri_Y124}  Polaronic transport can also result in this type mid-IR absorption.  Figure~\ref{sigma1_lowT} shows low temperature infrared data for the 100 nm film on STO.  Virtually no temperature dependence of this feature is observed down to 40 K.  This behaviour allows us to rule out the presence of small polarons, which would require a strong temperature dependence of the mid-IR absorption.\cite{polaron_Tdep}  Large polaron absorption spectra, on the other hand, are expected to be temperature independent and asymmetric,\cite{large_polarons} in agreement with our data.  The peak position near 0.3 eV is also consistent with large polarons, as it is below the 0.7 eV cut off predicted by Fr\"{o}hlich coupling for transition metal oxides~\cite{Dimitri_rmp}.  It has also been suggested that this type of absorption could be due to incoherent motion of carriers coupled to the spin degree of freedom. \cite{Tokura_jpsj} However, the lack of temperature dependence in our data makes this an unlikely scenario.   Finally, given that this is the only part of the spectra where differences between the three films are observed, it is also possible that feature A is the result of secondary phase scattering.

We now focus our attention on the far-IR part of the spectra in Fig.\ref{sigma1_roomT}.  Optical measurements are ideal for probing the kinetic energy of electrons, given that this quantity is proportional to the area under the Drude part of the optical conductivity, $K_{exp}\propto\int_0^\Omega\sigma_{1}(\omega)d\omega$.\cite{millis}  Comparison of the experimental electron kinetic energy to that obtained from band structure calculations, $K_{band}$, provides a means for classifying materials according to the strength of the electronic correlations in the system.\cite{millis2, mumtaz_nat_phys}  The ratio $K_{exp}/K_{band}$ is expected to be close to unity for itinerant electron systems and will become suppressed as strong interactions come into play.  Band structure calculations for LNO predict a finite density of states at the Fermi level and a plasma frequency $\omega_{p}=3.7 eV$.  Data in Fig. \ref{sigma1_roomT} however, show no obvious sign of a Drude contribution to the optical conductivity.  The plasma frequency obtained by integrating $\sigma_{1}(\omega)$ up to 1500 cm$^{-1}$ is 1 eV for the film on LAO and 0.8 eV for the films on STO.  This results in $K_{exp}/K_{band}\approx0.05-0.07$, which is quite low, even compared to other correlated metals.\cite{mumtaz_nat_phys}  In this context, our data present robust evidence of very strong electronic correlations in LNO.

In contrast with other lanthanide rare earth nickelates,\cite{NNO_MIT, SNO_PNO_optics} the strength of correlations in LNO does not appear to be tuned by temperature.  The low temperature data in Fig.\ref{sigma1_lowT} show no significant temperature dependence, except for the peak at 300 cm$^{-1}$, which we discuss below.    We note that, even in the realm of strongly correlated systems, the low energy optical conductivity spectra we have obtained is unusual.  Correlated metals typically show coherent (Drude) and incoherent contributions of similar strength.  In the case of LNO, the incoherent part of the optical conductivity (feature A) is much stronger than the coherent one.  

We now discuss the strong far-IR resonance shown in Figs. \ref{sigma1_roomT} and ~\ref{sigma1_lowT}.  According to group theory, three infrared-active phonon modes are expected in this range for a cubic perovskite.  The broad peak centred at 300 cm$^{-1}$ is composed of much more than three oscillators.  One possibility is that the modes are split and broadened due to the rhombohedral distortion of the LNO lattice.  At 40~K, the strength of the main peak is increased by about 20$\%$ and a small peak is resolved at 170 cm$^{-1}$ (Fig. \ref{sigma1_lowT}) .  This, along with some sharpening of the spectra at low temperature, is suggestive of absorption due to phonons.  The effective ionic charge obtained from this resonance by means of the equation $\int \sigma _{1}(\omega) d\omega = \dfrac{\pi n Z^{2}}{2m*} $ is \textit{Z}=2.5.  While this value is somewhat high, it is not outside the range of known effective ionic charges in oxides.\cite{Tokura_jpsj,conductivity_of_oxides}  In nickelates, \textit{Z}=1.9 has been obtained for La$_2$NiO$_{4+\delta}$.\cite{la2nio4}  As seen in Fig. \ref{sigma1_roomT}, no thickness dependence of this feature is observed, with the 100 nm and 200 nm films on STO showing almost identical far-IR spectra.   The film on LAO shows optical conductivity about 50$\%$ higher than the films on STO, consistent with resistivity data.  The shape of the resonance however, is remarkably similar even though LAO has a rhombohedral perovskite crystal structure and STO is a cubic perovskite.     

In light of our findings, the prediction of superconductivity in LaNiO$_3$/LaAlO$_3$ superlattices might need to be reexamined.  The authors of Ref. \onlinecite{superconductivity_theory} explain that bulk LNO has one electron in two degenerate \textit{e$_{g}$} orbitals.  In the proposed superlattice, the insulating LAO layers would confine the 3\textit{z}$^2$ - 1 orbital in the \textit{z} direction, possibly removing it from the Fermi energy.  The result is a single electron in a two-dimensional \textit{x}$^2$ - \textit{y}$^2$ band, reminiscent of the band structure of cuprates.  This model is presumably sensitive to the orbital structure of bulk LNO near the Fermi energy, in particular the \textit{e$_{g}$} levels.  While no conclusions about the possible orbital structure of LNO/LAO superlattices can be drawn from our work, we have clearly shown that the low frequency optical conductivity of LNO is in disagreement with band structure calculations.  As a result, we propose that the Ni \textit{e$_{g}$} orbitals may not be accurately described by current theoretical models.   Further work on fully strained ultra-thin films is needed to determine the effects of strain on the band structure and electronic properties of LNO, which will play a crucial role in the properties of LNO superlattices.

A recent x-ray photoemission study of LNO has found a dispersive \textit{e$_{g}$} band crossing the Fermi energy, in apparent disagreement with our data.\cite{photoemission}  A similar discrepancy between ARPES and infrared optical data has been reported for layered manganites and can be explained by the presence of a pseudogap at the Fermi energy.\cite{manganites, manganites_no_drude}  It is also important to note that LNO is a polar material in which electronic reconstruction at the surface is a likely possibility.\cite{polar_reconstruction}  In this case, one can expect a dramatic difference between optical measurements probing the entire depth of the films and ARPES, a surface probe. 

\section{Conclusions}

We have obtained optical conductivity of various LNO thin films over a wide temperature and energy range.  Above 1 eV, our data show several interband transition in good agreement with band structure calculations.  At lower frequencies however, significant discrepancies with theory are evident.  Despite band theory predictions of a finite density of states at the Fermi level, no Drude peak is present in our data.  We claim that this is evidence of strong correlations in LNO, which must be taken into account by theoretical models used in the design of oxide heterostructures.  We have shown that the strength of the electronic correlations is not controlled by temperature.    

\section{Aknowledgements}

The authors thank A. J. Millis, S. J. Moon, and G. Sawatzky for useful discussion.  J.C. was supported by DOD-ARO under the grant No. 0402-17291 and NSF grant No. DMR-0747808.

\end{document}